\documentclass[a4paper]{article}
\usepackage{amsmath}
\usepackage{amssymb}
\usepackage[a4paper,includeheadfoot]{geometry}

\begin{document}

\title{Renormalization:\ The observable-state model part II}
\author{Juan Sebasti\'{a}n Ardenghi \\
%EndAName
IFISUR,  Departamento de F\'{\i}sica (UNS-CONICET)  \and Mario
Castagnino \\
%EndAName
Institutos de F\'{\i}sica de Rosario y de Astronom\'{\i}a y F\'{\i}sica del
Espacio.\\
Casilla de Correos 67, sucursal 28, 1428 Buenos Aires, Argentina.}
\maketitle

\begin{abstract}
The purpose of this work is to rewrite the generating functional of $\phi
^{4}$ theory for the $n=0$ and $n=4$ correlation functions as the inner
product of a state with an observable, as we did in a previous work, for the
two-points correlation function. The observables are defined through the
external sources and the states are defined through the correlation function
itself. In this sense, the divergences of Quantum Field Theory (QFT) appear
in the reduced state by taking the partial trace of the state with respect
to the internal vertices that appear in the perturbation expansion. From
this viewpoint, the renormalization can be substituted by applying a
projector on the internal quantum state. The advantage of this new insight
is that we can obtain finite contributions to the correlation functions
without introducing counterterms in the Lagrangian or by manipulating
complex divergent quantities.
\end{abstract}

\section{Introduction}

This paper, as its predecessor, develops the perturbation expansion of any
correlation function in terms of the mean values of some observables in
particular states as we did in \cite{PR1}.\footnote{%
This idea has been called "the observable-state model".} In fact, our
formalism produce unphysical infinities in the form of $\left[ \delta (0)%
\right] ^{k}$ that will be represented in a dimensional regularization
scheme by the poles $\frac{1}{\epsilon ^{k}}$, where $\epsilon =d-4$ and $d$
is the space-time dimension.\footnote{%
The equivalence between $\delta (0)$ and $\frac{1}{\epsilon }$ can be found
in Quantum Field Theory textbooks, like \cite{Greiner}, page 352, below
eq.(11.55). In appendix A we show how to obtain this equivalence in a formal
way.} These infinities arise because the quantum state associated to the
internal vertices of the perturbation expansion has a diagonal part in the
coordinate basis. In \cite{PR1} we have shown that we can simply disregard
these unphysical infinities applying a projection operator on the quantum
states. The finite results found coincide with those of the usual
renormalized QFT in several models (and we will present more coincidences in
this and forthcoming papers). In this sense, it seems that throwing away the
unphysical infinities due to the short-distance behavior through the
projector is, after all, a good method. These ideas agree with those
introduced in \cite{W} (vol. 1, page 499): QFT yields divergent integrals
\textquotedblleft but these infinities cancel when we express all the
parameters of the theory in renormalized\ quantities, such as the masses and
the charges that we actually measure\textquotedblright . Moreover, it also
coincides with \cite{Folland}, since we believe that the process of
subtracting infinities is really a matter of subtracting the irrelevant
effect of the \textquotedblleft perhaps poorly understood physics at high
energy or short scale to obtain the meaningful physics at the scales
actually studied in the laboratory\textquotedblright\ (\cite{Folland}, page
254). In this sense, the constraining is done by neglecting the physics of
high energy or short scale.

\subsection{List of sections}

The paper is organized as follows:

In section 2 we will explicitly show how to define the observables and
states in a general way and the projection procedure.

In section 3 we will show how to describe the $n=0$ correlation function in $%
\phi ^{4}$ theory using the observables and states.

In section 4 we show in a similar way how to handle the observable-state
model for the $n=4$ correlation function. In particular, we show how the
renormalization group of the coupling constant arises.

In section 5 we show how to obtain the renormalization group equations for
the mass and the coupling constant using the finite contribution of the
correlation function obtained by application of the projector on the quantum
state.

In section 6 we briefly discuss the conceptual meaning of the reduced state
and partial traces and its relation with the non-physical virtual particles
and we introduce some general ideas of the observable-state model.

In section 7 we present the conclusions.

The appendix A shows the relation between the Dirac delta and the pole
parameter representation of the dimensional regularization. The appendix B
shows how to obtain the relation between the vacuum energy and the space
volume. Finally, in appendix C we analyze the properties of the projector
that gives the finite contribution in each correlation function.

\section{Observables and states in quantum field theory: the main idea}

Let us recall the main idea of the observable-state model of paper \cite{PR1}%
, that can be considered as the first part of this paper, and that will be
used in this section. The starting point is some (symmetric) $n$-point
functions $\tau ^{(n)}(x_{1},...,x_{n})$ (like Feynman or Euclidean
functions), and its corresponding generating functional (\cite{Haag}, eq.
(II.2.21), \cite{Brown}, eq. (3.2.11)). Then, the main equation reads:

\begin{equation}
iZ\left[ J\right] =\underset{n=0}{\overset{\infty }{\sum }}\underset{p=0}{%
\overset{\infty }{\sum }}\frac{i^{n}}{n!}\frac{i^{p}}{p!}\int \left\langle
\Omega _{0}\left\vert T\phi _{0}(x_{1})...\phi _{0}(x_{n})\mathcal{L}%
_{I}^{0}(y_{1})...\mathcal{L}_{I}^{0}(y_{p})\right\vert \Omega
_{0}\right\rangle J(x_{1})...J(x_{n})\underset{i=1}{\overset{n}{\prod }}%
d^{4}x_{i}\underset{i=1}{\overset{p}{\prod }}d^{4}y_{i}  \label{ideas7}
\end{equation}%
where $y_{i}$ are the internal vertices of the perturbation expansion and $%
\mathcal{L}_{I}^{0}(y_{p})$ is the Lagrangian interaction density (see
eq.(II.2.33) of \cite{Haag}).

This last equation will be our starting point, we will write $Z[J]$ as an
mean value of an observable defined through the $J(x_{n})$ sources in a
quantum state defined by the correlation function $\left\langle \Omega
_{0}\left\vert T\phi (x_{1})...\phi (x_{n})\mathcal{L}_{I}^{0}(y_{1})...%
\mathcal{L}_{I}^{0}(y_{p})\right\vert \Omega _{0}\right\rangle $.\footnote{%
In some sense, these observables will be the particle detector (see \cite%
{Thnew}, page 6, below eq.(2.6)).} This procedure will be done for each
correlation function of $n$ external points.

Using dimensional regularization (see \cite{TW}) we can write the
one-particle irreducible contribution to the correlation function such that
(see \cite{critical} for $\phi ^{4}$ theory):

\begin{equation}
\int \left\langle \Omega _{0}\left\vert T\phi (x_{1})...\phi (x_{n})%
\mathcal{L}_{I}^{0}(y_{1})...\mathcal{L}_{I}^{0}(y_{p})\right\vert \Omega
_{0}\right\rangle \underset{i=1}{\overset{p}{\prod }}%
d^{4}y_{i}=f_{0}^{(n)}(x_{1},...,x_{n})\overset{+\infty }{\underset{l=-L(n,p)%
}{\sum }}\beta _{l}^{(n,p)}(m_{0}^{2},\mu )\epsilon ^{l}  \label{la1}
\end{equation}%
where $f_{0}^{(n)}$ is some function of the external points, $\beta
_{l}^{(n,p)}(m_{0}^{2},\mu )$ are some coefficients of the dimensional
regularization that depends on the external momentum, the mass factor $\mu $
used to keep the coupling constant dimensionless and the mass of the field $%
m_{0}$. The parameter $\epsilon $ is $\epsilon =d-4$, where $d$ is the
dimension of space-time. The sum in $l$ starts at $-L(n,p)$ where $L(n,p)$
is the number of loops at order $p$ in the correlation functions of $n$
external points (see Appendix A, eq.(A6) of \cite{PR1}). The functions $%
f_{0}^{(n)}$ and $L(n,p)$ are very simple in the case of $\phi ^{4}$ theory,
for example

\begin{itemize}
\item \textbf{n=0}
\end{itemize}

\begin{equation}
f_{0}^{(0)}=1\text{\ \ \ \ , \ \ \ }L(0,p)=p+1\text{ }  \label{la2}
\end{equation}

\begin{itemize}
\item \textbf{n=2}
\end{itemize}

\begin{equation}
f_{0}^{(2)}=\int \frac{d^{4}p}{(2\pi )^{4}}\frac{e^{-ip(x_{1}-x_{2})}}{%
(p^{2}-m_{0}^{2})^{2}}\text{\ \ \ \ , \ \ \ \ }L(2,p)=p  \label{la3}
\end{equation}

\begin{itemize}
\item \textbf{n=4}
\end{itemize}

\begin{equation}
f_{0}^{(4)}=\int \frac{d^{4}p}{(2\pi )^{4}}\frac{e^{-ip(x_{1}-x_{4})}}{%
p^{2}-m_{0}^{2}}\int \frac{d^{4}q}{(2\pi )^{4}}\frac{e^{-iq(x_{2}-x_{4})}}{%
q^{2}-m_{0}^{2}}\int \frac{d^{4}l}{(2\pi )^{4}}\frac{e^{-il(x_{3}-x_{4})}}{%
(l^{2}-m_{0}^{2})((p+q+l)^{2}-m_{0}^{2})}\text{\ \ \ \ , \ \ \ \ }L(4,p)=p-1
\label{la4}
\end{equation}%
In general%
\begin{equation}
f_{0}^{(n)}=\overset{n}{\underset{i=1}{\prod }}\int \frac{d^{4}p_{i}}{(2\pi
)^{4}}\frac{e^{-ip_{i}x_{i}}}{p_{i}^{2}-m_{0}^{2}}\delta (\underset{j=1}{%
\overset{n}{\sum }}p_{j}^{2})\text{, \ \ \ \ \ \ \ \ \ \ \ \ \ \ \ \ \ \ \ \ 
}L(n,p)=p-\frac{n}{2}+1  \label{la5}
\end{equation}%
Inserting eq.(\ref{la1}) in eq.(\ref{ideas7}) we obtain\footnote{%
The infinite sum in the $l$ index in eq.(\ref{la1}) can be truncated in $l=0$%
, because the remaining terms are proportional to $\epsilon ^{l}$ and the
final result must be computed by taking the $\epsilon \rightarrow 0$ limit.
In this sense, what concern us is the principal part plus the constant term
of the Laurent serie with poles $d-4$.}%
\begin{equation}
iZ\left[ J\right] =\underset{n=0}{\overset{\infty }{\sum }}\underset{p=0}{%
\overset{\infty }{\sum }}\frac{i^{n}}{n!}\frac{i^{p}}{p!}\overset{0}{%
\underset{l=-L(n,p)}{\sum }}\beta _{l}^{(n,p)}\epsilon ^{l}\int
f_{0}^{(n)}(x_{1},...,x_{n})J(x_{1})...J(x_{n})\underset{i=1}{\overset{n}{%
\prod }}d^{4}x_{i}  \label{la6}
\end{equation}%
The observable-state model consist in the assumption that the generating
functional of last equation can be rewritten as a mean value of the
following observable

\begin{equation}
O^{(n,p)}=O_{ext}^{(n)}\otimes I_{int}^{(p)}  \label{lala0}
\end{equation}%
in the following quantum state%
\begin{equation}
\rho ^{(n,p)}=\rho _{ext}^{(n)}\otimes \rho _{int}^{(n,p)}  \label{lala0.1}
\end{equation}%
where $O_{ext}^{(n)}$ is some observable that acts on the external
coordinates $x_{i}$ and $I_{int}^{(p)}$ is the identity operator that acts
on the internal vertices due to the perturbation expansion. In a similar
way, $\rho _{ext}^{(n)}$ is the quantum state of the external part and $\rho
_{int}^{(n,p)}$ is the quantum state of the internal part.

Then, the mean value of $O^{(n,p)}$ in $\rho ^{(n,p)}$ reads%
\begin{equation}
Tr(\rho ^{(n,p)}O^{(n,p)})=Tr(\rho _{ext}^{(n)}O_{ext}^{(n)})Tr(\rho
_{int}^{(n,p)})  \label{lala0.2}
\end{equation}%
Using last equation, the generating functional of eq.(\ref{la6}) can be
written as%
\begin{equation}
iZ\left[ J\right] =\underset{n=0}{\overset{\infty }{\sum }}\underset{p=0}{%
\overset{\infty }{\sum }}\frac{i^{n}}{n!}\frac{i^{p}}{p!}Tr(\rho
^{(n,p)}O^{(n,p)})=\underset{n=0}{\overset{\infty }{\sum }}\underset{p=0}{%
\overset{\infty }{\sum }}\frac{i^{n}}{n!}\frac{i^{p}}{p!}Tr(\rho
_{int}^{(n,p)})Tr(\rho _{ext}^{(n)}O_{ext}^{(n)})  \label{lala0.3}
\end{equation}%
where%
\begin{equation}
\rho _{ext}^{(n)}=\int f_{0}^{(n)}(x_{1},...,x_{n})\left\vert x_{1},...,x_{%
\frac{n}{2}}\right\rangle \left\langle x_{\frac{n}{2}+1},...,x_{n}\right%
\vert \underset{i=1}{\overset{n}{\prod }}d^{4}x_{i}  \label{lala0.4}
\end{equation}%
and%
\begin{equation}
O_{ext}^{(n)}=\int J(x_{1})...J(x_{n})\left\vert x_{1},...,x_{\frac{n}{2}%
}\right\rangle \left\langle x_{\frac{n}{2}+1},...,x_{n}\right\vert \underset{%
i=1}{\overset{n}{\prod }}d^{4}x_{i}  \label{lala0.5}
\end{equation}%
In turn%
\begin{equation}
Tr(\rho _{int}^{(n,p)})=\overset{+\infty }{\underset{l=-L(n,p)}{\sum }}\beta
_{l}^{(n,p)}\epsilon ^{l}  \label{lala0.6}
\end{equation}%
which implies that the divergences of the quantum field theory are the
consequence of taking the trace of the internal quantum state $\rho
_{int}^{(n,p)}$. This point is relevant; because the trace of an operator is
an invariant quantity, this means that it is the same in different basis.
This implies that if we want to obtain a finite contribution $\beta
_{0}^{(n,p)}$, we must apply a non-unitary transformation on $\rho
_{int}^{(n,p)}$ that changes its trace, i.e., we must project to another $%
\rho _{int}$.

\subsection{Internal quantum state}

To define the internal quantum state we will just recall some considerations
(see Section VI in \cite{PR1}): the algebra of observables $\mathcal{O}$ is
represented by $^{\ast }-$algebra $\mathcal{A}$ of self-adjoint elements and
states are represented by functionals on $\mathcal{O}$, that is, by elements
of the dual space $\mathcal{O}^{\prime }$, $\rho \in \mathcal{O}^{\prime }$.
We will construct a $C^{\ast }-$algebra of operators defined in terms of
elements with the property $Tr(A^{\ast }A)<\infty $. As it is well known, a $%
C^{\ast }-$algebra can be represented in a Hilbert space $\mathcal{H}$ (GNS\
theorem)\footnote{%
Gelfand, Naimark and Segal \cite{GNS}.} and, in this particular case $%
\mathcal{O=\mathcal{O}}^{\prime }$; therefore$\mathcal{\ \mathcal{O}}$ and $%
\mathcal{O}^{\prime }$ are represented by $\mathcal{H\otimes \mathcal{H}}$
that will be called $\mathcal{N}$, the Liouville space.

As we are interested in the diagonal and non-diagonal elements of a matrix
state we can define a sub algebra of $\mathcal{\mathcal{\mathcal{N}}}$, that
can be called a van Hove algebra \cite{van hove} since such a structure
appears in his work as:%
\begin{equation}
\mathcal{\mathcal{\mathcal{N}}}_{vh}\mathcal{=N}_{S}\oplus \mathcal{N}%
_{R}\subset \mathcal{\mathcal{\mathcal{N}}}  \label{k}
\end{equation}%
where the vector space $\mathcal{N}_{R}$ is the space of operators with $%
O(x)=0$ and $O(x,x^{\prime })$ is a regular function. Moreover $\mathcal{O=N}%
_{vhS}$ is the space of selfadjoint operators of $\mathcal{\mathcal{\mathcal{%
N}}}_{vh},$ which can be constructed in such a way it could be dense in $%
\mathcal{N}_{S}$ (because any distribution can be approximated by regular
functions) (for the details see \cite{PR1}, Section II.B and Section VI).
Therefore essentially the introduced restriction is the minimal possible
coarse-graining. Now the $\oplus $ is a direct sum because $\mathcal{N}_{S}$
contains the factor $\delta (x-x^{\prime })$ and $\mathcal{N}_{R}$ contains
just regular functions and a kernel cannot be both a $\delta $ and a regular
function. Moreover, as our observables must be self-adjoint, the space of
observables must be 
\begin{equation}
\mathcal{O=N}_{vhS}\mathcal{=N}_{S}\oplus \mathcal{N}_{R}\subset \mathcal{N}
\label{l}
\end{equation}

The states must be considered as linear functionals over the space $\mathcal{%
O}$ ($\mathcal{O}^{\prime}$ the dual of space $\mathcal{O}$):

\begin{equation}
\mathcal{O}^{\prime }\mathcal{=N}_{vhS}^{\prime }\mathcal{=N}_{S}^{\prime
}\oplus \mathcal{N}_{R}^{\prime }\subset \mathcal{N}^{\prime }  \label{m}
\end{equation}%
The set of these generalized states is the convex set $\mathcal{S\subset O}%
^{\prime }$.

Having this in mind, we can define the internal quantum state in the
following way%
\begin{eqnarray}
\rho _{int}^{(n,p)} &=&\int \underset{i=1}{\overset{L(n,p)}{\prod }}\left(
\rho _{D}^{(n,p,i)}(y_{i})\delta (y_{i}-w_{i})+\rho
_{ND}^{(n,p,i)}(y_{i},w_{i})\right)  \label{c9.1} \\
&&\left\vert y_{1},...,y_{L(n,p)}\right\rangle \left\langle
w_{1},..,w_{L(n,p)}\right\vert \underset{i=1}{\overset{L(n,p)}{\prod }}%
d^{4}y_{i}d^{4}w_{i}  \notag
\end{eqnarray}%
\bigskip

The trace reads (see Appendix B, eq.(\ref{delta7})):%
\begin{equation}
Tr(\rho _{int}^{(n,p)})=\underset{i=1}{\overset{L(n,p)}{\prod }}\left( 
\frac{\rho _{D}^{(n,p,i)}}{\pi \epsilon }+\rho _{ND}^{(n,p,i)}\right)
\label{c9.3}
\end{equation}%
where%
\begin{equation}
\rho _{D}^{(n,p,i)}=\int\limits_{{}}^{{}}\rho
_{D}^{(n,p,i)}(y_{i})d^{4}y_{i}\text{ \ \ \ \ \ \ \ \ }\rho
_{ND}^{(n,p,i)}=\int\limits_{{}}^{{}}\rho
_{ND}^{(n,p,i)}(y_{i},y_{i})d^{4}y_{i}  \label{c9.4}
\end{equation}%
We can see from last equation that $\rho _{D}^{(n,p,i)}$ and $\rho
_{ND}^{(n,p,i)}$ are merely normalization factors. Eq.(\ref{c9.3}) can be
written as%
\begin{equation}
Tr(\rho _{int}^{(n,p)})=\underset{l=-L(n,p)}{\overset{0}{\sum }}\gamma
_{l}^{(n,p)}\epsilon ^{l}  \label{c9.50}
\end{equation}%
where%
\begin{equation}
\gamma _{0}^{(n,p)}=\underset{i=1}{\overset{L(n,p)}{\prod }}\rho
_{ND}^{(n,p,i)}\text{, ... \ , \ \ }\gamma _{L(n,p)}^{(n,p)}=\frac{1}{\pi
^{L(n,p)}}\underset{i=1}{\overset{L(n,p)}{\prod }}\rho _{D}^{(n,p,i)}
\label{c9.501}
\end{equation}

All the terms $\gamma _{l}^{(n,p)}$ with $l>0$ that are multiplied by $%
\epsilon ^{l}$ contain at least one $\rho _{D}^{(n,p,i)}$, that is, the
diagonal part of the state of the $\dot{i}-$internal quantum system. In
particular, we can make the following equality%
\begin{equation}
\beta _{l}^{(n,p)}=\gamma _{l}^{(n,p)}  \label{c9.7}
\end{equation}

In this sense, the coefficients obtained by the dimensional regularization
can be associated with the products of the diagonal and non-diagonal parts
of the internal quantum state. In particular, the coefficient that is not
multiplied by a $\epsilon $ is $\gamma _{0}^{(n,p)}$ which depends
exclusively on the non-diagonal quantum state.

\subsection{Projection over the finite contribution}

As we saw in eq.(\ref{c9.50}) and eq.(\ref{c9.501}), the finite result
exclusively depends on the non-diagonal quantum state, so we can construct a
projector that projects over the non-diagonal quantum state. This projector
reads\footnote{%
Is not difficult to show that it is a projector: linearity implies that $\Pi
(a+b)=\Pi (a)+\Pi (b)$, then, if $\Pi (a)=a-G$, then, $\Pi ^{2}(a)=\Pi
(a-G)=\Pi (a)-\Pi (G)$, but $\Pi (G)=G-G=0$, then $\Pi ^{2}(a)=\Pi (a)$.}%
\begin{eqnarray}
\Pi _{p}(\rho _{int}^{(n,p)})=\rho _{int}^{(n,p)}-\int \rho
_{D}^{(n,p,1)}(y_{1})\rho _{D}^{(n,p,2)}(y_{2})...\rho
_{D}^{(n,p,L(n,p))}(y_{L(n,p)})\left\vert y_{1},...,y_{L(n,p)}\right\rangle
\left\langle y_{1},..,y_{L(n,p)}\right\vert \underset{i=1}{\overset{L(n,p)}{%
\prod }}d^{4}y_{i}  \label{finite1} \\
+\int \rho _{D}^{(n,p,1)}(y_{1})\rho _{D}^{(n,p,2)}(y_{2})...\rho
_{D}^{(n,p,L(n,p)-1)}(y_{L(n,p)-1})\rho
_{ND}^{(n,p,L(n,p))}(y_{L(n,p)},w_{L(n,p)})  \notag \\
\left\vert y_{1},...,y_{L(n,p)}\right\rangle \left\langle
y_{1},..,w_{L(n,p)}\right\vert d^{4}w_{L(n,p)}\underset{i=1}{\overset{%
L(n,p)-1}{\prod }}d^{4}y_{i}+...+\int \rho _{D}^{(n,p,1)}(y_{1})\rho
_{ND}^{(n,p,2)}(y_{2},w_{2})...\rho
_{ND}^{(n,p,L(n,p))}(y_{L(n,p)},w_{L(n,p)})  \notag \\
\left\vert y_{1},...,y_{L(n,p)}\right\rangle \left\langle
y_{1},..,w_{L(n,p)}\right\vert d^{4}y_{1}\underset{i=2}{\overset{L(n,p)}{%
\prod }}d^{4}y_{i}d^{4}w_{i})  \notag
\end{eqnarray}%
The projection procedure consists in the subtraction of the part of the\
state that contains at least one internal diagonal quantum state. This
projector acting on the state $\rho ^{(n,p)}$ yields%
\begin{equation}
\Pi _{p}(\rho _{int}^{(n,p)})=\int \underset{i=1}{\overset{L(n,p)}{\prod }}%
\rho _{ND}^{(n,p,i)}(y_{i},w_{i})\left\vert
y_{1},...,y_{L(n,p)}\right\rangle \left\langle
w_{1},..,w_{L(n,p)}\right\vert \underset{i=1}{\overset{L(n,p)}{\prod }}%
d^{4}y_{i}d^{4}w_{i}  \label{finite2}
\end{equation}%
Then, using the equivalence of eq.(\ref{c9.7}), the mean value of $O^{(n,p)}$
in the state $\Pi _{p}(\rho ^{(n,p)})$ reads:%
\begin{equation}
Tr(\Pi _{p}(\rho ^{(n,p)})O^{(n,p)})=\beta _{0}^{(n,p)}\int
f_{0}^{(n)}(x_{1},...,x_{n})O_{ext}^{(n)}\left( x_{1},...,x_{n}\right) 
\underset{i=1}{\overset{n}{\prod }}d^{4}x_{i}  \label{finite3}
\end{equation}%
where $O_{ext}^{(n)}\left( x_{1},...,x_{n}\right) =J(x_{1})...J(x_{n})$ (see
eq.(\ref{lala0.5})). Multiplying by $\frac{i^{p}}{p!}$ and summing in $p$ we
obtain\footnote{%
The factor $\frac{i^{p}}{p!}$ is introduced for later convenience, but its
meaning could be that in the observable-state model, the quantum state is
invariant under an exchange of internal vertices.}%
\begin{equation}
Tr(\rho ^{(n)}O_{ext}^{(n)})=\overset{+\infty }{\underset{p=0}{\sum }}\frac{%
i^{p}}{p!}Tr(\Pi _{p}(\rho ^{(n,p)})O^{(n,p)})=\overset{+\infty }{\underset{%
p=0}{\sum }}\frac{i^{p}}{p!}\beta _{0}^{(n,p)}\int
f_{0}^{(n)}(x_{1},...,x_{n})O_{ext}^{(n)}\left( x_{1},...,x_{n}\right) 
\underset{i=1}{\overset{n}{\prod }}d^{4}x_{i}  \label{finite3.1}
\end{equation}%
where%
\begin{equation}
\rho ^{(n)}=\left( \overset{+\infty }{\underset{p=0}{\sum }}\frac{i^{p}}{p!}%
\beta _{0}^{(n,p)}\right) \rho _{ext}^{(n)}  \label{finite3.2}
\end{equation}%
where $\frac{i^{p}}{p!}\beta _{0}^{(n,p)}$ is the coefficient of the quantum
state $\rho _{ext}^{(n)}$.

In this way, we can eliminate all the divergences of the observable-state
model by the application of the projector over a well defined Hilbert
subspace. This formalism has been applied to the two-point correlation
function for $\phi ^{4}$ theory (see \cite{PR1}) and the idea of this work
is to apply it to $n=0$ and $n=4$ correlation function of external points.
In appendix C we briefly show the relation between the projector and the $R-$%
operation of the \textit{BPHZ} subtraction method in QFT.

\section{Examples:\ $\protect\phi ^{4}$ theory, $n=0$}

In this section we will briefly study the vacuum amplitude for the $\phi
^{4} $ theory. When there are interactions, the vacuum amplitude reads (see 
\cite{PS}, page 87):%
\begin{equation}
\left\langle \Omega \right\vert \Omega \rangle =\left( \left\vert
\left\langle \Omega _{0}\right\vert \Omega \rangle \right\vert
^{2}e^{-iE_{0}2T}\right) ^{-1}\left\langle \Omega _{0}\left\vert \exp
(-i\int\limits_{-T}^{T}dtH_{I}(t))\right\vert \Omega _{0}\right\rangle
\label{va1}
\end{equation}%
where $\left\vert \Omega \right\rangle $ is the vacuum vector for the
interacting theory, $\left\vert \Omega _{0}\right\rangle $ is the vacuum
vector for the free theory, $E_{0}=\left\langle \Omega \left\vert
H\right\vert \Omega \right\rangle $ is the energy of the vacuum state of the
interacting theory, $H$ is the full Hamiltonian $H=H_{0}+H_{I}$ where $H_{I}$
is the interacting Hamiltonian and $2T$ is the time interval where the
process occurs. The brackets in eq.(\ref{va1}) can be written in terms of
the perturbation expansion in the coupling constant $\lambda _{0}$:%
\begin{gather}
\left\langle \Omega _{0}\left\vert \exp
(-i\int\limits_{-T}^{T}dtH_{I}(t))\right\vert \Omega _{0}\right\rangle
=1+\left( -i\lambda _{0}\right) \int\limits_{{}}^{{}}d^{4}y_{1}\left\langle
\Omega _{0}\left\vert \phi ^{4}(y_{1})\right\vert \Omega _{0}\right\rangle +
\label{va2} \\
\left( -i\lambda _{0}\right)
^{2}\int\limits_{{}}^{{}}d^{4}y_{1}d^{4}y_{2}\left\langle \Omega
_{0}\left\vert \phi ^{4}(y_{1})\phi ^{4}(y_{2})\right\vert \Omega
_{0}\right\rangle +...+\left( -i\lambda _{0}\right)
^{p}\int\limits_{{}}^{{}}d^{4}y_{1}...d^{4}y_{p}\left\langle \Omega
_{0}\left\vert \phi ^{4}(y_{1})...\phi ^{4}(y_{p})\right\vert \Omega
_{0}\right\rangle +...  \notag
\end{gather}%
The structure of the vacuum amplitude in terms of the perturbation expansion
can be obtained, to do so we will consider the first order in the
perturbation expansion. We just recall that we will compute the connected
diagrams and not the products of them.

The first order $p=1$ reads:%
\begin{equation}
\left( -i\lambda _{0}\right) \int\limits_{{}}^{{}}d^{4}y_{1}\left\langle
\Omega _{0}\left\vert \phi ^{4}(y_{1})\right\vert \Omega _{0}\right\rangle
=i\lambda _{0}\left[ \Delta (0)\right] ^{2}\int%
\limits_{{}}^{{}}d^{4}y_{1}=i\lambda _{0}\left[ \Delta (0)\right] ^{2}2TV
\label{va4}
\end{equation}%
where $V$ is the volume of space and $\Delta (0)\,$is the Feynman propagator
of a scalar field. Using dimensional regularization, eq.(\ref{va4}) reads%
\begin{equation}
(-i\lambda _{0})\int\limits_{{}}^{{}}d^{4}y_{1}\left\langle \Omega
_{0}\left\vert \phi ^{4}(y_{1})\right\vert \Omega _{0}\right\rangle
=i\lambda _{0}2TV\left( \frac{\beta _{2}^{(0,1)}}{\epsilon ^{2}}+\frac{\beta
_{1}^{(0,1)}}{\epsilon }+\beta _{0}^{(0,1)}\right)   \label{va5}
\end{equation}%
where the coefficients $\beta _{i}^{(0,1)}$ are some constants that can be
obtained from the regularized propagator $\Delta (0)$ and depends on a mass
factor $\mu $ that is introduced to keep the coupling constant
dimensionless, this is, we must replace $\lambda _{0}$ by $\lambda
_{0}\left( \mu ^{-\epsilon }\right) $.\footnote{%
Is not difficult to show that the coupling constant has dimension $\left[
\lambda _{0}\right] =$mass$^{4-d}$ where $d$ is the dimension of space-time
(see \cite{PS}, page 322). Then, the mass factor $\mu ^{-(4-d)}$ multiplied
to $\lambda _{0}$ mantains the new coupling constant dimensionless. A
dimensionless coupling constant is necessary because it is the parameter we
use to apply the perturbation expansion.} The first superscript $0$ in $%
\beta $ refers to the number of external points and the second superscript $1
$ refers to the order in the perturbation expansion. The subscript refers to
the power of the $\epsilon =d-4$ factor, where $d$ is the dimension of
space-time. Using eq. (A.44)\ of Appendix A.4 of \cite{PS}, page 807), the
coefficients $\beta _{k}^{(0,1)}$ reads%
\begin{gather}
\beta _{2}^{(0,1)}=\frac{m_{0}^{4}}{64\pi ^{4}}  \label{va5.1} \\
\beta _{1}^{(0,1)}=\frac{m_{0}^{4}}{64\pi ^{4}}\left( \gamma -1+\ln \left( 
\frac{m_{0}^{2}}{4\pi \mu }\right) \right)   \notag \\
\beta _{0}^{(0,1)}=\frac{m_{0}^{4}}{24\cdot 64\pi ^{4}}\left( 18-24\gamma
+12\gamma ^{2}+\pi ^{2}+12\left( \ln ^{2}(m_{0}^{2})-\ln ^{2}(4\pi )+\ln
^{2}(\mu )\right) +24(1-\gamma +\ln (4\pi ))\ln (\frac{4\pi \mu }{m_{0}^{2}}%
)\right)   \notag
\end{gather}%
The second order $p=2$ in the perturbation expansion has three terms, where
two of them are connected:%
\begin{gather}
\left( -i\lambda _{0}\right)
^{2}\int\limits_{{}}^{{}}d^{4}y_{1}d^{4}y_{2}\left\langle \Omega
_{0}\left\vert \phi ^{4}(y_{1})\phi ^{4}(y_{2})\right\vert \Omega
_{0}\right\rangle =\left( -i\lambda _{0}\right) ^{2}\left[ \Delta (0)\right]
^{2}\int\limits_{{}}^{{}}d^{4}y_{1}d^{4}y_{2}\left[ \Delta (y_{1}-y_{2})%
\right] ^{2}  \label{va6} \\
+\left( -i\lambda _{0}\right) ^{2}\int\limits_{{}}^{{}}d^{4}y_{1}d^{4}y_{2}
\left[ \Delta (y_{1}-y_{2}\right] ^{4}  \notag
\end{gather}

It can be shown that the following orders for the connected Feynman diagrams
in the perturbation expansion can be accommodated following eq. (\ref{va5}):%
\footnote{%
The general solution showed in eq.(\ref{va13}) can be traced to general
results which appears in the dimensional regularization scheme ((see \cite%
{critical}, page 103-130 and \cite{ashok}, page 686)).}%
\begin{equation}
\left( -i\lambda _{0}\right)
^{p}\int\limits_{{}}^{{}}d^{4}y_{1}...d^{4}y_{p}\left\langle \Omega
_{0}\left\vert \phi ^{4}(y_{1})...\phi ^{4}(y_{p})\right\vert \Omega
_{0}\right\rangle =\overset{p+1}{\underset{j=0}{\sum }}\left( -i\lambda
_{0}\right) ^{p}i^{2p}(2TV)\frac{\beta _{j}^{(0,p)}}{\epsilon ^{j}}
\label{va13}
\end{equation}%
where $i^{2p}$ comes from $2p$ propagators that can be obtain from the
vacuum expectation values of the $4p$ quantum fields. If we want to compute
eq.(\ref{va1}) we must consider the non-connected Feynman diagrams that can
be constructed by multiplying the connected ones. For example, for the
second order $p=2$ we can obtain the non-connected Feynman diagram by
multiplying by itself the first order $p=1$. This procedure can be done for
all the orders, in particular, to obtain the non-connected Feynman diagrams
at order $p$ we must multiply all the lowest orders where the sum of them
gives $p$. If we call the result of eq.(\ref{va13}) as $f(p)$, then, the sum
of the connected diagrams and the non-connected diagrams reads%
\begin{equation}
\underset{k=0}{\overset{+\infty }{\sum }}\frac{1}{k!}\left( \underset{p=1}{%
\overset{+\infty }{\sum }}f(p)\right) ^{k}=f(1)+f(2)+...+f(p)+...+\frac{1}{2!%
}\left( f(1)+f(2)+...\right) \left( f(1)+f(2)+...\right) +...  \label{va13.1}
\end{equation}%
the factor $\frac{1}{k!}$ is introduced to avoid double counting, for
example $f(i)f(j)$ and $f(j)f(i)$. With this result, we can proceed to
evaluate eq.(\ref{va1}):%
\begin{equation}
\left\vert \left\langle \Omega _{0}\right\vert \Omega \rangle \right\vert
^{2}e^{-iE_{0}2T}=\underset{k=0}{\overset{+\infty }{\sum }}\frac{1}{k!}%
\left( \underset{p=1}{\overset{+\infty }{\sum }}\overset{p+1}{\underset{j=0}{%
\sum }}\left( -i\lambda _{0}\right) ^{p}i^{2p}(2TV)\frac{\beta _{j}^{(0,p)}}{%
\epsilon ^{j}}\right) ^{k}  \label{va14}
\end{equation}%
where we have put $\left\langle \Omega \right\vert \Omega \rangle =1$ and we
have introduced the result of eq.(\ref{va13}) in $f(p)$. The projection
procedure will be given by only keeping the $j=0$ term in eq. (\ref{va14})
as we will show in the following section. We then have:%
\begin{equation}
\left\vert \left\langle \Omega _{0}\right\vert \Omega \rangle \right\vert
^{2}e^{-iE_{0}2T}=\underset{k=0}{\overset{+\infty }{\sum }}\frac{\left(
2TV\right) ^{k}}{k!}\left( \underset{p=1}{\overset{+\infty }{\sum }}\left(
-i\lambda _{0}\right) ^{p}i^{2p}\beta _{0}^{(0,p)}\right) ^{k}
\label{va15.1}
\end{equation}

In Appendix A we show how to obtain the relation between the vacuum energy $%
E_{0}$ and the volume of space $V$ in a formal way. This result has no
direct relation with the aim of this work, but is a contribution to the
observable-state model.

\subsection{The observable-state model for $n=0$ in $\protect\phi ^{4}$}

Now we can apply this mathematical structure to the case of vacuum bubbles
in $\phi ^{4}$ theory, where we can use eq.(\ref{c9.1}) in the case $n=0$,
then:%
\begin{equation}
\rho ^{(0,p)}=\rho _{int}^{(0,p)}=\int\limits_{{}}^{{}}\underset{i=1}{%
\overset{p+1}{\prod }}(\rho _{D}^{(0,p,i)}(y_{i})\delta (y_{i}-w_{i})+\rho
_{ND}^{(0,p,i)}(y_{i},w_{i}))\left\vert y_{1},...,y_{p+1}\right\rangle
\left\langle w_{1},...,w_{p+1}\right\vert \underset{i=1}{\overset{p+1}{\prod 
}}d^{4}y_{i}d^{4}w_{i}  \label{va22}
\end{equation}%
where $\rho _{D}^{(0,p,i)}$ and $\rho _{ND}^{(0,p,i)}$ are some regular
functions. The trace $Tr(\rho _{int}^{(0,p)})$ reads%
\begin{equation}
Tr(\rho _{int}^{(0,p)})=\underset{l=-(p+1)}{\overset{0}{\sum }}\gamma
_{l}^{(0,p)}\epsilon ^{l}  \label{va24}
\end{equation}%
where in particular%
\begin{equation}
\gamma _{0}^{(0,p)}=\underset{i=1}{\overset{p+1}{\prod }}\rho
_{ND}^{(0,p,i)}\text{ \ , \ ... \ , \ \ }\gamma _{p+1}^{(0,p)}=\frac{1}{\pi
^{p+1}}\underset{i=1}{\overset{p+1}{\prod }}\rho _{D}^{(0,p,i)}
\label{va25.2}
\end{equation}%
and the remaining coefficients $\gamma _{l}^{(0,p)}$ with $p+1>l>1$ contains
at least one $\rho _{D}$.

Comparing eq.(\ref{va25.2}) with eq.(\ref{va13}) we can see that the
coefficients $\gamma _{l}^{(0,p)}$ read%
\begin{equation}
\gamma _{l}^{(0,p)}=\left( -i\lambda _{0}\right) ^{p}i^{2p}(2TV)\beta
_{l}^{(0,p)}  \label{va25.3}
\end{equation}%
In the first order in the perturbation expansion, using eq.(\ref{va24}) and
eq.(\ref{va25.2}) we have%
\begin{gather}
\underset{l=0}{\overset{2}{\sum }}\gamma _{l}^{(0,p)}\epsilon ^{l}=\gamma
_{0}^{(0,2)}+\gamma _{1}^{(0,2)}\epsilon ^{-1}+\gamma _{2}^{(0,2)}\epsilon
^{-2}  \label{va25.4} \\
=\rho _{ND}^{(0,2,1)}\rho _{ND}^{(0,2,2)}+(\rho _{D}^{(0,2,1)}\rho
_{ND}^{(0,2,2)}+\rho _{D}^{(0,2,2)}\rho _{ND}^{(0,2,1)})\epsilon ^{-1}+\rho
_{D}^{(0,2,1)}\rho _{D}^{(0,2,2)}\epsilon ^{-2}  \notag
\end{gather}%
Using eq.(\ref{va5.1}) and eq.(\ref{va25.3}) we have that :%
\begin{gather}
\rho _{D}^{(0,2,1)}\rho _{D}^{(0,2,2)}=\frac{i\lambda _{0}2TVm_{0}^{2}}{%
64\pi ^{4}}  \label{va25.5} \\
\rho _{D}^{(0,2,1)}\rho _{ND}^{(0,2,2)}+\rho _{D}^{(0,2,2)}\rho
_{ND}^{(0,2,1)}=\frac{i\lambda _{0}2TVm_{0}^{2}}{64\pi ^{4}}\left( -1+\gamma
+\ln \left( \frac{m_{0}^{2}}{4\pi \mu }\right) \right)  \notag \\
\rho _{ND}^{(0,2,1)}\rho _{ND}^{(0,2,2)}=\frac{i\lambda _{0}2TVm_{0}^{4}}{%
24\cdot 64\pi ^{4}}\left( 18-24\gamma +12\gamma ^{2}+\pi ^{2}+12\left( \ln
^{2}(m_{0}^{2})-\ln ^{2}(4\pi )+\ln ^{2}(\mu )\right) +24(1-\gamma +\ln
(4\pi ))\ln (\frac{4\pi \mu }{m_{0}^{2}})\right)  \notag
\end{gather}

This implies that the diagonal and nondiagonal quantum states are not well
determined. In this case, we have four unknown quantities and three
equations. As we saw in Section 2, the finite contribution for the
correlation function comes from the non-diagonal quantum states, so the
indetermination can be translate to an arbitrary election of one of the
non-diagonal quantum state. The indetermination will grow up with the order
of the perturbation expansion; in fact, at order $p$ we will have $p$
diagonal states and $p$ non-diagonal states, so we have $2p$ unknown
quantities, but we have $p+1$ equations, so the indetermination grows like $%
2p-p-1=p-1$. In general, for the correlation function of $n$ external points
we will have $2L(n,p)$ unknown quantities and $L+1$ equations, so the
indetermination will grow as $2L-L-1=L-1$.

The finite contribution of eq. (\ref{va24}) can be obtained by the
application of the projector on the quantum state of eq.(\ref{va22}):%
\begin{eqnarray}
\Pi _{p}(\rho _{int}^{(0,p)})=\rho _{int}^{(0,p)}-\int \rho
_{D}^{(0,p,1)}(y_{1})\rho _{D}^{(0,p,2)}(y_{2})...\rho
_{D}^{(0,p,p+1)}(y_{p+1})\left\vert y_{1},...,y_{p+1}\right\rangle
\left\langle y_{1},..,y_{p+1}\right\vert \underset{i=1}{\overset{p+1}{\prod }%
}d^{4}y_{i}+  \label{va30.1} \\
\int \rho _{D}^{(0,p,1)}(y_{1})\rho _{D}^{(0,p,2)}(y_{2})...\rho
_{ND}^{(0,p,p+1)}(y_{p+1},w_{p+1})\left\vert y_{1},...,y_{p+1}\right\rangle
\left\langle y_{1},..,w_{p+1}\right\vert d^{4}w_{p+1}\underset{i=1}{\overset{%
p}{\prod }}d^{4}y_{i}+...  \notag \\
...+\int \rho _{D}^{(0,p,1)}(y_{1})\rho _{ND}^{(0,p,2)}(y_{2},w_{2})...\rho
_{ND}^{(0,p,p+1)}(y_{p+1},w_{p+1})\left\vert y_{1},...,y_{p+1}\right\rangle
\left\langle y_{1},..,w_{p+1}\right\vert d^{4}y_{1}\underset{i=2}{\overset{%
p+1}{\prod }}d^{4}y_{i}d^{4}w_{i}  \notag
\end{eqnarray}%
This projector eliminates all the diagonal parts of the quantum state. Then,
the trace with the projected state reads%
\begin{equation}
Tr(\Pi _{p}(\rho _{int}^{(0,p)}))=\beta _{0}^{(0,p)}  \label{va30.3}
\end{equation}%
Adding all the orders in the perturbation expansion we finally obtain%
\begin{equation}
Tr(\Pi (\rho _{int}^{(0)}))=1+\underset{p=1}{\overset{+\infty }{\sum }}%
\left( -i\lambda _{0}\right) ^{p}i^{2p}(2TV)\beta _{0}^{(0,p)}
\label{va30.4}
\end{equation}%
Then, multiplying the non-connected Feynman diagrams, we obtain eq.(\ref%
{va15.1}).

In the case of no external points, the renormalization is a normalization of
the quantum state itself. In the observable-state model, this normalization
is explicit, because the projection changes the trace of the quantum state
(see eqs.(\ref{va24}) and (\ref{va30.3})). From this point of view, the
renormalization is a change of the norm of the quantum state by a
projection, in a similar manner in which the projection postulate occurs in
non-relativistic quantum mechanics.

\section{Example: $\protect\phi ^{4}$ theory, $n=4$}

The four-point correlation function, when there are interactions, reads%
\begin{eqnarray}
\left\langle \Omega \left\vert \phi (x_{1})\phi (x_{2})\phi (x_{3})\phi
(x_{4})\right\vert \Omega \right\rangle =\left\langle \Omega _{0}\left\vert
\phi (x_{1})\phi (x_{2})\phi (x_{3})\phi (x_{4})\right\vert \Omega
_{0}\right\rangle +  \label{n4.1} \\
\left( -i\lambda _{0}\right) \int\limits_{{}}^{{}}\left\langle \Omega
_{0}\left\vert \phi (x_{1})\phi (x_{2})\phi (x_{3})\phi (x_{4})\phi
^{4}(y_{1})\right\vert \Omega _{0}\right\rangle d^{4}y_{1}+  \notag \\
+\left( -i\lambda _{0}\right) ^{2}\int\limits_{{}}^{{}}\left\langle \Omega
_{0}\left\vert \phi (x_{1})\phi (x_{2})\phi (x_{3})\phi (x_{4})\phi
^{4}(y_{1})\phi ^{4}(y_{2})\right\vert \Omega _{0}\right\rangle
d^{4}y_{1}d^{4}y_{2}+...  \notag \\
+...+\left( -i\lambda _{0}\right) ^{p}\int\limits_{{}}^{{}}\left\langle
\Omega _{0}\left\vert \phi (x_{1})\phi (x_{2})\phi (x_{3})\phi (x_{4})\phi
^{4}(y_{1})\phi ^{4}(y_{2})\right\vert \Omega _{0}\right\rangle \underset{i=1%
}{\overset{p}{\prod }}d^{4}y_{p}+...  \notag
\end{eqnarray}%
The first term of the last equation reads%
\begin{equation}
\left\langle \Omega _{0}\left\vert \phi (x_{1})\phi (x_{2})\phi (x_{3})\phi
(x_{4})\right\vert \Omega _{0}\right\rangle =\Delta (x_{1}-x_{2})\Delta
(x_{3}-x_{4})+\Delta (x_{1}-x_{3})\Delta (x_{2}-x_{4})+\Delta
(x_{1}-x_{4})\Delta (x_{2}-x_{3})  \label{n4.2}
\end{equation}%
where $\Delta (x-y)$ is the scalar propagator. This term does not contribute
to the scattering amplitude because it describes a trivial process where the
initial and final states are identical.

The first order in the perturbation expansion reads%
\begin{eqnarray}
\left( -i\lambda _{0}\right) \int\limits_{{}}^{{}}\left\langle \Omega
_{0}\left\vert \phi (x_{1})\phi (x_{2})\phi (x_{3})\phi (x_{4})\phi
^{4}(y_{1})\right\vert \Omega _{0}\right\rangle
d^{4}y_{1}=f_{0}^{(4)}(x_{1},x_{2},x_{3},x_{4})  \label{n4.3} \\
=\left( -i\lambda _{0}\right) \int \frac{d^{4}p}{(2\pi )^{4}}\frac{%
ie^{-ip(x_{1}-x_{4})}}{p^{2}-m_{0}^{2}}\int \frac{d^{4}q}{(2\pi )^{4}}\frac{%
ie^{-iq(x_{2}-x_{4})}}{q^{2}-m_{0}^{2}}\int \frac{d^{4}l}{(2\pi )^{4}}\frac{%
ie^{-il(x_{3}-x_{4})}}{(l^{2}-m_{0}^{2})}\frac{i}{((p+q-l)^{2}-m_{0}^{2})} 
\notag
\end{eqnarray}%
In this case, the first order does not have any loops.

The second order in the perturbation expansion reads%
\begin{eqnarray}
\left( -i\lambda _{0}\right) ^{2}\int\limits_{{}}^{{}}\left\langle \Omega
_{0}\left\vert \phi (x_{1})\phi (x_{2})\phi (x_{3})\phi (x_{4})\phi
^{4}(y_{1})\phi ^{4}(y_{2})\right\vert \Omega _{0}\right\rangle
d^{4}y_{1}d^{4}y_{2}=  \label{n4.4} \\
f_{0}^{(4)}(x_{1},x_{2},x_{3},x_{4})\lambda _{0}^{2}\int \frac{d^{4}r}{%
(2\pi )^{4}}\frac{1}{(r^{2}-m_{0}^{2})((p+q-r)^{2}-m_{0}^{2})}=  \notag \\
f_{0}^{(4)}(x_{1},x_{2},x_{3},x_{4})\lambda _{0}^{2}\left( \frac{\beta
_{1}^{(4,2)}}{\epsilon }+\beta _{0}^{(4,2)}\right)   \notag
\end{eqnarray}%
where $\beta _{1}^{(4,2)}$ and $\beta _{0}^{(4,2)}$ read (see \cite{Ramond},
page 120-122 or in eq.(4.4.16)):%
\begin{eqnarray}
\beta _{1}^{(4,2)}=\frac{1}{32\pi ^{2}}  \label{n.4.4.1} \\
\beta _{0}^{(4,2)}=\frac{1}{2}\frac{3}{32\pi ^{2}}\left( \ln (\mu
^{2})-\gamma +2+\ln (\frac{4\pi \mu ^{2}}{m_{0}^{2}})-\frac{1}{3}\underset{%
z=s,t,u}{\sum }\sqrt{1+\frac{4m_{0}^{2}}{z}}\ln \left( \frac{\sqrt{1+\frac{%
4m_{0}^{2}}{z}}+1}{\sqrt{1+\frac{4m_{0}^{2}}{z}}-1}\right) \right)   \notag
\end{eqnarray}%
where $s$, $t\,$and $u$ are Mandelstam variables $s=(p_{1}+p_{2})^{2}$, $%
t=(p_{1}+p_{3})^{2}$ and $u=(p_{1}+p_{4})^{2}$ and $\frac{1}{2}$ is the
symmetry factor and the $\mu $ factor appears by changing the coupling
constant $\lambda _{0}$ to $\lambda _{0}\mu ^{-\epsilon }$. Is not difficult
to show that the higher orders in the perturbation expansion obey the
following rule%
\begin{gather}
\left( -i\lambda _{0}\right) ^{p}\int\limits_{{}}^{{}}\left\langle \Omega
_{0}\left\vert \phi (x_{1})\phi (x_{2})\phi (x_{3})\phi (x_{4})\phi
^{4}(y_{1})...\phi ^{4}(y_{p})\right\vert \Omega _{0}\right\rangle \underset{%
i=1}{\overset{p}{\prod }}d^{4}y_{p}=  \label{n4.5} \\
f_{0}^{(4)}(x_{1},x_{2},x_{3},x_{4})\underset{l=0}{\overset{p-1}{\sum }}%
\frac{\left( -i\lambda _{0}\right) ^{p}i^{2+2p}\beta _{l}^{(4,p)}}{\epsilon
^{l}}  \notag
\end{gather}%
where $p-1$ is the number of loops in the case of $\phi ^{4}$ theory with
four external points.

Following the idea of our work, we will apply the observable-state model to
the four-point correlation function.

\subsection{The observable-state model for $n=4$ in $\protect\phi ^{4}$
theory}

The state and the observable reads%
\begin{eqnarray}
\rho ^{(4,p)} &=&\int f_{0}^{(4)}(x_{1},x_{2},x_{3},x_{4})\underset{i=1}{%
\overset{p-1}{\prod }}\left( \rho _{D}^{(4,p,i)}(y_{i})\delta
(y_{i}-w_{i})+\rho _{ND}^{(4,p,i)}(y_{i},w_{i})\right)  \label{os4.1} \\
&&\left\vert x_{1},x_{2},y_{1}...,y_{p-1}\right\rangle \left\langle
x_{3},x_{4},w_{1},..,w_{p-1}\right\vert \underset{i=1}{\overset{4}{\prod }}%
d^{4}x_{i}\underset{i=1}{\overset{p-1}{\prod }}d^{4}y_{i}d^{4}w_{i}  \notag
\end{eqnarray}

\begin{equation}
O^{(4,p)}=\int J(x_{1})J(x_{2})J(x_{3})J(x_{4})\underset{i=1}{\overset{p-1}{%
\prod }}\delta (y_{i}-w_{i})\left\vert
x_{1},x_{2},y_{1},...,y_{p-1}\right\rangle \left\langle
x_{3},x_{4},w_{1},...,w_{p-1}\right\vert \underset{i=1}{\overset{4}{\prod }}%
d^{4}x_{i}\underset{i=1}{\overset{p-1}{\prod }}d^{4}y_{i}d^{4}w_{i}
\label{os4.2}
\end{equation}%
Then, the trace reads%
\begin{equation}
Tr(\rho ^{(4,p)}O^{(4,p)})=\overset{p-1}{\underset{l=0}{\sum }}\frac{\gamma
_{l}^{(4,p)}}{\epsilon ^{l}}\int
f_{0}^{(4)}(x_{1},x_{2},x_{3},x_{4})J(x_{1})J(x_{2})J(x_{3})J(x_{4})\underset%
{i=1}{\overset{4}{\prod }}d^{4}x_{i}  \label{os4.3}
\end{equation}%
where%
\begin{equation}
\gamma _{l}^{(4,p)}=\left( -i\lambda _{0}\right) ^{p}i^{2+2p}\beta
_{l}^{(4,p)}  \label{os4.5}
\end{equation}%
In particular%
\begin{equation}
\gamma _{0}^{(4,p)}=\underset{i=1}{\overset{p-1}{\prod }}\rho
_{ND}^{(4,p,i)}\text{ \ , \ ... \ , \ \ }\gamma _{p-1}^{(4,p)}=\frac{1}{\pi
^{p-1}}\underset{i=1}{\overset{p-1}{\prod }}\rho _{D}^{(4,p,i)}
\label{os4.6}
\end{equation}%
For the order $p=2$, using eq. (\ref{n.4.4.1})\ and eq.(\ref{os4.5}), the $%
\gamma _{l}^{(4,2,1)}$ coefficients read%
\begin{eqnarray}
\gamma _{1}^{(4,2)}=\rho _{D}^{(4,2,1)}=\frac{\lambda _{0}^{2}}{32\pi ^{2}}
\label{os4.6.1} \\
\gamma _{0}^{(4,2)}=\rho _{ND}^{(4,2,1)}=\frac{\lambda _{0}^{2}}{32\pi ^{2}}%
\left( -\frac{1}{2}\ln (\mu )-\gamma +2+\ln (\frac{4\pi \mu }{m_{0}^{2}})-%
\sqrt{1+\frac{4m_{0}^{2}}{(p+q)^{2}}}\ln \left( \frac{\sqrt{1+\frac{%
4m_{0}^{2}}{(p+q)^{2}}}+1}{\sqrt{1+\frac{4m_{0}^{2}}{(p+q)^{2}}}-1}\right)
\right)  \notag
\end{eqnarray}%
The projector over the finite contribution reads%
\begin{eqnarray}
\Pi _{p}(\rho _{int}^{(4,p)})=\rho _{int}^{(4,p)}-\int \rho
_{D}^{(4,p,1)}(y_{1})\rho _{D}^{(4,p,2)}(y_{2})...\rho
_{D}^{(4,p,p-1)}(y_{p-1})\left\vert y_{1},...,y_{p-1}\right\rangle
\left\langle y_{1},..,y_{p-1}\right\vert \underset{i=1}{\overset{p-1}{\prod }%
}d^{4}y_{i}+  \label{os4.7} \\
\int \rho _{D}^{(4,p,1)}(y_{1})\rho _{D}^{(4,p,2)}(y_{2})...\rho
_{ND}^{(4,p,p-1)}(y_{p-1},w_{p-1})\left\vert y_{1},...,y_{p-1}\right\rangle
\left\langle y_{1},..,w_{p-1}\right\vert d^{4}w_{p-1}\underset{i=1}{\overset{%
p-2}{\prod }}d^{4}y_{i}+...  \notag \\
...+\int \rho _{D}^{(4,p,1)}(y_{1})\rho _{ND}^{(4,p,2)}(y_{2},w_{2})...\rho
_{ND}^{(4,p,p-1)}(y_{p-1},w_{p-1})\left\vert y_{1},...,y_{p-1}\right\rangle
\left\langle y_{1},..,w_{p-1}\right\vert d^{4}y_{1}\underset{i=2}{\overset{%
p-1}{\prod }}d^{4}y_{i}d^{4}w_{i}  \notag
\end{eqnarray}%
Then, the trace of the observable in the projected state reads%
\begin{equation}
Tr(\Pi _{p}\rho ^{(4,p)}O^{(4,p)})=\gamma _{0}^{(4,p)}\int \rho
_{ext}^{(4,1)}(x_{1},x_{2},x_{3},x_{4})J(x_{1})J(x_{2})J(x_{3})J(x_{4})%
\underset{i=1}{\overset{4}{\prod }}d^{4}x_{i}  \label{os4.8}
\end{equation}%
Summing all the perturbation expansion terms we obtain%
\begin{equation}
Tr(\Pi \rho ^{(4)}O^{(4)})=\int
f_{0}^{(4)}(x_{1},x_{2},x_{3},x_{4})J(x_{1})J(x_{2})J(x_{3})J(x_{4})\underset%
{i=1}{\overset{4}{\prod }}d^{4}x_{i}=\underset{p=0}{\overset{+\infty }{\sum }%
}\left( -i\lambda _{0}\right) ^{p}i^{2+2p}\beta _{0}^{(4,p)}  \label{os4.9}
\end{equation}%
where we have replaced $\gamma _{0}^{(4,p)}$ by $\left( -i\lambda
_{0}\right) ^{p}i^{2+2p}\beta _{0}^{(4,p)}$ (see eq.(\ref{os4.5})).

\subsection{Renormalization of $\protect\lambda $}

We can proceed by summing the perturbation expansion, but without taking
account the $p=0$ order, because it describes a trivial process in which the
initial and final states are identical. Only fully connected diagrams
contribute to the scattering amplitude. Then%
\begin{equation}
\left\langle \Omega \left\vert \phi (x_{1})\phi (x_{2})\phi (x_{3})\phi
(x_{4})\right\vert \Omega \right\rangle =f_{0}^{(4)}(x_{1},x_{2},x_{3},x_{4})%
\underset{p=1}{\overset{+\infty }{\sum }}\underset{l=0}{\overset{p-1}{\sum }}%
\frac{\left( -i\lambda _{0}\right) ^{p}i^{2+2p}\beta _{l}^{(4,p)}}{\epsilon
^{l}}  \label{renor1}
\end{equation}%
We can then put $x_{4}=0$ and take the Fourier transform on both sides of
last equation:%
\begin{eqnarray}
\int
d^{4}x_{1}d^{4}x_{2}d^{4}x_{3}e^{-ipx_{1}}e^{-iqx_{2}}e^{-ilx_{3}}\left%
\langle \Omega \left\vert \phi (x_{1})\phi (x_{2})\phi (x_{3})\phi
(0)\right\vert \Omega \right\rangle =  \label{renor2} \\
\frac{1}{(p^{2}-m_{0}^{2})}\frac{1}{(q^{2}-m_{0}^{2})}\frac{1}{%
(l^{2}-m_{0}^{2})}\frac{1}{((p+q-l)^{2}-m_{0}^{2})}\underset{p=1}{\overset{%
+\infty }{\sum }}\underset{l=0}{\overset{p-1}{\sum }}\frac{\left( -i\lambda
_{0}\right) ^{p}i^{2+2p}\beta _{l}^{(4,p)}}{\epsilon ^{l}}  \notag
\end{eqnarray}

If we remove the propagators of the external lines we obtain the four point
proper vertex $\Gamma ^{(4)}$. We can write $i\Gamma ^{(4)}(0)=\lambda $,
this is, the renormalized coupling constant is equal to the magnitude of the
scattering amplitude at zero momentum (see \cite{PS}, page 325). But from
dimensional regularization we know that the coupling constant depends on the
mass factor $\mu $, so in the most general case, $i\Gamma ^{(4)}=\lambda
(\mu )$, then%
\begin{equation}
i\lambda (\mu )=\underset{p=1}{\overset{+\infty }{\sum }}\left( -i\lambda
_{0}\right) ^{p}i^{2+2p}\underset{l=0}{\overset{p-1}{\sum }}\frac{\beta
_{l}^{(4,p)}}{\epsilon ^{l}}  \label{renor3}
\end{equation}%
where $\beta _{l}^{(4,p)}$ depends on $\mu $ and the external momentum. The
last equation is identical to eq.(2.3.b) of \cite{thooft}. Once
renormalized, we must only keep the $l=0$ term, then%
\begin{equation}
i\lambda (\mu )=\underset{p=1}{\overset{+\infty }{\sum }}\left( -i\lambda
_{0}\right) ^{p}i^{2+2p}\beta _{0}^{(4,p)}  \label{renor4}
\end{equation}%
In terms of the observable-state model, this reads (see eq.(\ref{os4.5})):%
\begin{equation}
i\lambda (\mu )=\underset{p=1}{\overset{+\infty }{\sum }}\gamma _{0}^{(4,p)}
\label{renor5}
\end{equation}%
In this sense, the non-diagonal functions of the quantum state of eq.(\ref%
{os4.1}), that is, the renormalized coupling constant.

\section{The renormalization group}

In this last section we will see how the renormalization group arises in the
context of the observable-state model. As we see in \cite{PR1} and this
paper, the $n=2$ and $n=4$ correlation functions give the mass and coupling
constant renormalization. Those equations read (see eq.(B18) of \cite{PR1}
and eq.(\ref{renor4}) of this paper):\footnote{%
In the following equations we will restore the Planck constant $\hbar $ for
later convenience.}\ 
\begin{equation}
m^{2}=m_{0}^{2}+\underset{p=1}{\overset{+\infty }{\sum }}\left( -i\lambda
_{0}\right) ^{p}i^{1+2p}\hbar ^{p}\beta _{0}^{(2,p)}(m_{0}^{2},\mu
)=m_{0}^{2}-\lambda _{0}\hbar \beta _{0}^{(2,1)}(m_{0}^{2},\mu )+...
\label{rg1}
\end{equation}%
\begin{equation}
\lambda =\lambda _{0}+\underset{p=2}{\overset{+\infty }{\sum }}\left(
-i\lambda _{0}\right) ^{p}i^{2+2p}\hbar ^{p-1}\beta
_{0}^{(4,p)}(m_{0}^{2},\mu )=\lambda _{0}+\lambda _{0}^{2}\hbar \beta
_{0}^{(4,2)}(m_{0}^{2},\mu )+...  \label{rg2}
\end{equation}%
In the other side, since $m_{0}^{2}$ and $\lambda _{0}$ do not depend on $%
\mu $ in the absence of loop correction, we have:%
\begin{equation}
\frac{dm_{0}^{2}}{d\mu }=O(\hbar )\text{ \ \ \ \ \ \ , \ \ \ \ \ }\frac{%
d\lambda _{0}}{d\mu }=O(\hbar )  \label{rg2.2}
\end{equation}

The renormalization group can be obtained by imposing the fact that the
dressed mass $m^{2}$ and $\lambda $ do not depend on $\mu $, this is, $\frac{%
dm^{2}}{d\mu }=0$ and $\frac{d\lambda }{d\mu }=0$. Using the chain rule in
eq.(\ref{rg1}), we have for $m^{2}$:%
\begin{equation}
\frac{dm^{2}}{d\mu }=\frac{\partial m^{2}}{\partial m_{0}^{2}}\frac{%
dm_{0}^{2}}{d\mu }+\frac{\partial m^{2}}{\partial \lambda _{0}}\frac{%
d\lambda _{0}}{d\mu }+\frac{\partial m^{2}}{\partial \mu }=0  \label{rg3}
\end{equation}%
using eqs.(\ref{rg1}) and (\ref{rg2.2}), eq.\ref{rg3}) reads at order $\hbar 
$:%
\begin{equation}
\frac{dm_{0}^{2}}{d\mu }-\lambda _{0}\frac{\partial \beta _{0}^{(2,1)}}{%
\partial \mu }=0  \label{rg6}
\end{equation}%
From eq.(82) of \cite{PR1}%
\begin{equation}
\beta _{0}^{(2,1)}=\frac{m_{0}^{2}}{16\pi ^{2}}\left[ 1-\gamma +2\ln \left( 
\frac{4\pi \mu ^{2}}{m_{0}^{2}}\right) \right]  \label{rg7}
\end{equation}%
then%
\begin{equation}
\frac{\partial \beta _{0}^{(2,1)}}{\partial \mu }=\frac{m_{0}^{2}}{8\pi ^{2}}%
\frac{1}{\mu }  \label{rg8}
\end{equation}%
replacing eq.(\ref{rg8}) in eq.(\ref{rg6}) we obtain a differential equation
for $m_{0}^{2}$ at order $\hbar $:%
\begin{equation}
\frac{dm_{0}^{2}}{d\mu }=\lambda _{0}\frac{m_{0}^{2}}{8\pi ^{2}}\frac{1}{\mu 
}  \label{rg9}
\end{equation}%
we can solve it and obtain%
\begin{equation}
m_{0}^{2}=m_{S}^{2}(\frac{\mu }{\mu _{S}})^{\frac{\lambda _{0}}{8\pi ^{2}}}
\label{rg10}
\end{equation}%
where $m_{S}^{2}$ is the value of the mass when $\mu =$ $\mu _{S}$. This
result is in concordance with eq.(4.6.20) and eq.(4.6.22), page 142 of \cite%
{Ramond} at order $\hbar $. In a similar way, we can obtain the change of $%
\lambda _{0}$ in terms of $\mu $ at order $\hbar $. To do so, we must impose
that the dressed coupling constant do not depend on $\mu $:%
\begin{equation}
\frac{d\lambda }{d\mu }=\frac{\partial \lambda }{\partial m_{0}^{2}}\frac{%
dm_{0}^{2}}{d\mu }+\frac{\partial \lambda }{\partial \lambda _{0}}\frac{%
d\lambda _{0}}{d\mu }+\frac{\partial \lambda }{\partial \mu }=0  \label{rg11}
\end{equation}%
using eq. (\ref{rg2}) and eq.(\ref{rg2.2}), last equation reads at order $%
\hbar $:%
\begin{equation}
\lambda _{0}^{2}\frac{\partial \beta _{0}^{(4,2)}}{\partial \mu }-\frac{%
d\lambda _{0}}{d\mu }=0  \label{rg13}
\end{equation}%
Using the result of eq.(\ref{n.4.4.1}), last equation reads%
\begin{equation}
\frac{d\lambda _{0}}{d\mu }+\frac{3}{16\pi ^{2}}\frac{\lambda _{0}^{2}}{\mu }%
=0  \label{rg14}
\end{equation}%
The last equation can be solved with the following result:%
\begin{equation}
\lambda _{0}=\frac{\lambda _{S}}{1-\frac{3\lambda _{S}}{16\pi ^{2}}\ln (%
\frac{\mu }{\mu _{S}})}  \label{rg17}
\end{equation}%
which is identical to eq.(4.6.15), page 139 of \cite{Ramond}. This last
equation is the one-loop correction to the coupling constant that arises
from eq.(\ref{rg13}).\footnote{%
The power of the Planck constant counts the number of loops, so at order $%
O(\hbar )$, we obtain the one-loop correction (see \cite{ashok}, page 623).}

Thus, we can see that the projection method not only allow finite
perturbation expansions, but also, finite values {}{}that are consistent
with the results shown in textbooks and the renormalization group.

\section{Discussion}

The formalism introduced in Section 1 has a physical content which can be
traced to the decoherence formalism (see \cite{JPZ}, \cite{Deco}, \cite{C3}, 
\cite{Ordoniez}, \cite{Fortin}, \cite{Bonachon}) and to systems with
continuous spectrum (see \cite{castagnino}, \cite{C2}, \cite{Deco}, \cite{4'}%
,\cite{4-2}, \cite{C3}). The trace of the internal quantum state of eq.(\ref%
{lala0.6}) can be interpreted as a reduced state, since the observable is an
identity operator in the Hilbert space of the internal vertices.This has a
physical meaning. It is well known that the reduction of a state decreases
the information available to the observer about the composite system. In
this case, the reduction is done over the internal vertices where the
interaction occurs due to the perturbation expansion. In QFT, the particles
that are created in these vertices are virtual particles because they are
off-shell, that is, they do obey the conservation laws, but the propagators
must be integrated out, which implies that the momentum of the particle
associated with each internal propagator may not obey the mass-energy
relation $p_{\mu }p^{\mu }=m_{0}^{2}$. In this sense, the conceptual meaning
of the partial trace of the internal degrees of freedom is to neglect these
virtual non-physical particles. This is consistent with the experiments of
scattering because basically what is seen are the in and out states.
However, perturbation theory introduces off-shell intermediate states whose
existence depends on the uncertainty principle $\Delta E\Delta t\geq \frac{%
\hbar }{2}$. In turn, these give us an interpretation of this integration as
a reduction of the degrees of freedom of the theory. In the conventional
interpretation of this integration \textquotedblleft The integral $d^{4}z$
instruct us to sum over all points where this process can occur. This is
just the superposition principle of quantum mechanics:\ when a process can
happen in alternative ways, we add the amplitudes for each possible
way.\textquotedblright , (\cite{PS}, page 94). In our case, the integration
over the internal vertices reflects the fact that we are neglecting the
degrees of freedom of this virtual particles and what we finally obtain is a
reduced state which is divergent.

Summarizing, the main idea of this work is that in the $p$ order in the
perturbation expansion of any quantum field theory, we can define a quantum
state as%
\begin{equation}
\rho ^{(n,p)}=\rho _{ext}^{(n)}\otimes \rho _{int}^{(n,p)}  \label{tr1}
\end{equation}%
and an observable%
\begin{equation}
O^{(n,p)}=O_{ext}^{(n)}\otimes I_{int}^{(p)}  \label{tr2}
\end{equation}%
then the trace reads%
\begin{equation}
Tr(\rho ^{(n,p)}O^{(n,p)})=Tr(\rho _{int}^{(n,p)})Tr(\rho
_{ext}^{(n)}O_{ext}^{(n)})  \label{tr3}
\end{equation}%
The divergences of the quantum field theory occur in the trace of the
internal quantum state $Tr(\rho _{int}^{(n,p)})$. These divergences appear
because the internal quantum state contains diagonal functions multiplied by
Dirac deltas that cannot be avoided unless we remove the diagonal functions
by a projection. This is the only available transformation that can cure the
divergences, because the trace is an invariant quantity that does not depend
on the basis in which the state is written. The projector reads%
\begin{equation}
\Pi ^{(n,p)}=I_{ext}^{(n)}\otimes \Pi _{int}^{(n,p)}=I_{ext}^{(n)}\otimes
(\rho _{int}^{(n,p)}-\rho _{D}^{(n,p)})  \label{tr3.1}
\end{equation}%
where $\rho _{D}^{(n,p)}$ is the sum of all the states that has a diagonal
part of the quantum state $\rho _{int}^{(n,p)}$. Then, the trace of $\Pi
\rho ^{(n,p)}$ reads%
\begin{equation}
Tr(\Pi ^{(n,p)}\rho ^{(n,p)}O^{(n,p)})=\left( Tr(\rho
_{int}^{(n,p)})-Tr\left( \rho _{D}^{(n,p)}\right) \right) Tr(\rho
_{ext}^{(n)}O_{ext}^{(n)})  \label{tr3.2}
\end{equation}%
which is our finite desired physical contribution. Basically, the projection
is a translation of the quantum state by an amount given by the diagonal
state. In this work, the renormalization procedure is done by the projection
method, but without introducing counterterms, which in principle is much
more advantageous, because it can be applied to non-renormalizable theories,
like $\phi ^{6}$ in four space-time dimensions, or the quantum field theory
of a massless particle with spin $2$, such as gravitation. These two
theories will be worked out in future works.

\subsection{A general procedure}

Suppose we define the following projector that acts on the external quantum
state $\rho ^{(n)}$ of eq.(\ref{finite3.2}):%
\begin{equation}
\Pi _{0}^{(n)}=I_{1}\otimes I_{2}\otimes ...I_{n-1}\otimes \left\vert
0\right\rangle \left\langle 0\right\vert  \label{1}
\end{equation}%
where $\left\vert 0\right\rangle $ correspond to $x_{n}=0$. When we apply it
to $\rho ^{(n)}$ we obtain%
\begin{equation}
\rho ^{(n)}\Pi _{0}^{(n)}=\overset{+\infty }{\underset{p=0}{\sum }}\frac{%
i^{p}}{p!}\beta _{0}^{(n,p)}\int
f_{0}^{(n)}(x_{1},x_{2},...,x_{n-1},0)\left\vert x_{1},...,x_{\frac{n}{2}%
}\right\rangle \left\langle x_{\frac{n}{2}+1},...,x_{n-1},0\right\vert 
\underset{i=1}{\overset{n-1}{\prod }}d^{4}x_{i-1}  \label{2}
\end{equation}%
The trace with $O_{ext}^{(n)}$ reads%
\begin{equation}
Tr(\rho ^{(n)}\Pi _{0}^{(n)}O_{ext}^{(n)})=\overset{+\infty }{\underset{p=0}{%
\sum }}\frac{i^{p}}{p!}\beta _{0}^{(n,p)}\int
f_{0}^{(n)}(x_{1},x_{2},...,x_{n-1},0)J(x_{1})...J(x_{n-1})J(0)\underset{i=1}%
{\overset{n-1}{\prod }}d^{4}x_{i-1}  \label{3}
\end{equation}%
If we allow the currents to be plane waves:\footnote{%
This idea is in concordance with \cite{Thnew}, page 19, "For an ingoing
particle, we use a source function $J(x)$ whose Fourier components emit a
positive amount of energy $k_{0}$. For an out-going particle the source
emits a negative $k_{0}$."}%
\begin{equation}
J(x_{k})=e^{-ip_{k}x_{k}}  \label{4}
\end{equation}%
then, the trace reads%
\begin{equation}
Tr(\rho ^{(n)}\Pi _{0}^{(n)}\widetilde{O}_{ext}^{(n)})=\overset{+\infty }{%
\underset{p=0}{\sum }}\frac{i^{p}}{p!}\beta _{0}^{(n,p)}\mathcal{F}\left[
f_{0}^{(n)}(x_{1},x_{2},...,x_{n-1},0)\right] (k_{1},...k_{n-1})  \label{5}
\end{equation}%
where $\widetilde{O}_{ext}^{(n)}$ is the plane wave observable and $\mathcal{%
F}\left[ f\right] $ is the Fourier transform of the function $f$.

\begin{itemize}
\item \textbf{The mass shift }

In the case $n=2$:
\end{itemize}

\begin{equation}
f_{0}^{(2)}(x_{1},0)=\int \frac{d^{4}p}{(2\pi )^{4}}\frac{%
e^{-ip(x_{1}-x_{2})}}{(p^{2}-m_{0}^{2})^{2}}  \label{6}
\end{equation}%
then

\begin{equation}
Tr(\rho ^{(2)}\Pi _{0}^{(2)}{}\widetilde{O}_{ext}^{(2)})=\frac{1}{%
(p^{2}-m_{0}^{2})^{2}}\underset{p=1}{\overset{+\infty }{\sum }}\left(
-i\lambda _{0}\right) ^{p}\beta _{0}^{(2,p)}=\frac{M}{(p^{2}-m_{0}^{2})^{2}}
\label{7}
\end{equation}%
where $M=\underset{p=1}{\overset{+\infty }{\sum }}\left( -i\lambda
_{0}\right) ^{p}\beta _{0}^{(2,p)}$. Then, this equation implies that%
\begin{equation}
(p^{2}-m_{0}^{2})^{2}Tr(\rho ^{(2)}\Pi _{0}^{(2)}{}\widetilde{O}%
_{ext}^{(2)})=M  \label{7.1}
\end{equation}%
The mass renormalization is obtained by having in mind that the last
equation is the result of the one-particle irreducible diagrams.\footnote{%
A one-particle irreducible diagram is any diagram that cannot be split in
two by removing a single line.} The full contribution of the $n=2$
correlation function is equal to the following geometric series (see \cite%
{PS}, eq.(10.27), page 328):\ 
\begin{equation}
\int \left\langle \Omega \left\vert \phi (x_{1})\phi (x_{0})\right\vert
\Omega \right\rangle e^{-ipx_{1}}d^{4}x_{1}=\frac{1}{p^{2}-m_{0}^{2}}+\frac{M%
}{(p^{2}-m_{0}^{2})^{2}}+\frac{M^{2}}{(p^{2}-m_{0}^{2})^{3}}+...=\frac{1}{%
p^{2}-(m_{0}^{2}+M)}  \label{8.1}
\end{equation}%
On the other side, using eq.(using eq.(\ref{7})) we have%
\begin{equation}
\int \left\langle \Omega \left\vert \phi (x_{1})\phi (x_{0})\right\vert
\Omega \right\rangle e^{-ipx_{1}}d^{4}x_{1}=\frac{1}{%
p^{2}-m_{0}^{2}-(p^{2}-m_{0}^{2})^{2}Tr(\rho ^{(2)}\Pi _{0}^{(2)}{}%
\widetilde{O}_{ext}^{(2)})}  \label{8.2}
\end{equation}%
which implies the mass shift reads%
\begin{equation}
\Delta m=m^{2}-m_{0}^{2}=M=(p^{2}-m_{0}^{2})^{2}Tr(\rho ^{(2)}\Pi
_{0}^{(2)}{}\widetilde{O}_{ext}^{(2)})  \label{9.2}
\end{equation}

\begin{itemize}
\item \textbf{The coupling constant}

In the $n=4$ case
\end{itemize}

\begin{equation}
f_{0}^{(4)}(x_{1},x_{2},x_{3},0)=\int \frac{d^{4}p}{(2\pi )^{4}}\frac{%
e^{-ipx_{1}}}{p^{2}-m_{0}^{2}}\int \frac{d^{4}q}{(2\pi )^{4}}\frac{%
e^{-iqx_{2}}}{q^{2}-m_{0}^{2}}\int \frac{d^{4}l}{(2\pi )^{4}}\frac{%
e^{-ilx_{3}}}{(l^{2}-m_{0}^{2})}\frac{1}{((p+q-l)^{2}-m_{0}^{2})}  \label{10}
\end{equation}%
then

\begin{equation}
\left[
(l^{2}-m_{0}^{2})(p^{2}-m_{0}^{2})(q^{2}-m_{0}^{2})((p+q-l)^{2}-m_{0}^{2})%
\right] Tr(\rho ^{(4)}\Pi _{0}^{(4)}{}\widetilde{O}_{ext}^{(4)})=\underset{%
p=1}{\overset{+\infty }{\sum }}\left( -i\lambda _{0}\right) ^{p}\beta
_{0}^{(4,p)}=\lambda  \label{11}
\end{equation}%
which has the same structure of eq.(\ref{7.1}).

In a general way we can write%
\begin{equation}
Tr(\rho ^{(n)}\Pi _{0}^{(n)}{}\widetilde{O}_{ext}^{(n)})\int%
\limits_{{}}^{{}}\underset{i=1}{\overset{n}{\prod }}\left(
p_{i}^{2}-m_{0}^{2}\right) \delta (p_{n}-\underset{i=1}{\overset{n-1}{\sum }}%
p_{i})d^{4}p_{n}=C_{n}  \label{12}
\end{equation}%
where $C_{n}$ is the renormalized quantity.

This last equation is important, because it can be applied to
non-renormalizable theories. In \cite{K}, the renormalization group has been
generalized to Lagrangians of arbitrary form, in particular, to
non-renormalizables theories. The idea of this work and \cite{PR1} follows
the same line of thought because the observable-state model treats on equal
footing the non-renormalizable theories and the renormalizable ones.

\section{Conclusions}

The aim of this work was to extend the observable-state model in $\phi ^{4}$
theory to the $n=0$ and $n=4$ external points in the correlation function,
showing how to build a projector that eliminates all the divergences that
appear in the perturbation expansion. This procedure allows us to
renormalize the quantum field theory of $\phi ^{4}$ without introducing
counterterms in the Lagrangians. Besides this, we have shown how the
renormalization group arise in this context obtaining the same results as
the conventional renormalized QFT.

\section{Acknowledgment}

This paper was partially supported by grants of CONICET (Argentina National
Research Council), FONCYT (Argentina Agency for Science and Technology) and
the University of Buenos Aires.

\appendix

\section{The Dirac delta and the dimensional regularization poles}

To understand the relation between the Dirac delta and the poles of the
dimensional regularization we can use the following representation of the
Dirac delta (see \cite{gelfand}, page 35):\footnote{%
The relation between the Dirac delta and the \ dimensional regularization
pole in this appendix is introduced by formal mathematical operations, but
we must warm the reader that this development is not mathematically rigorous.%
}%
\begin{equation}
\delta (x)=\underset{\epsilon \rightarrow 0}{\lim }\frac{1}{\pi }\frac{%
\epsilon }{x^{2}+\epsilon ^{2}}  \label{delta1}
\end{equation}%
where $\epsilon $ is some parameter that tends to zero. In particular, we
can assume that this parameter is the pole parameter of the dimensional
regularization, that is, $\epsilon =d-4$.

Consider now for simplicity, the following quantum state%
\begin{equation}
\rho =\int [\rho _{D}(x)\delta (x-x^{\prime })+\rho _{ND}(x,x^{\prime
})]\left\vert x\right\rangle \left\langle x^{\prime }\right\vert
dxdx^{\prime }  \label{delta2}
\end{equation}%
replacing the representation of the Dirac delta of eq.(\ref{delta1}) in last
equation we obtain%
\begin{equation}
\rho =\frac{1}{\pi }\underset{\epsilon \rightarrow 0}{\lim }\int \rho
_{D}(x)\frac{\epsilon }{(x-x^{\prime })^{2}+\epsilon ^{2}}\left\vert
x\right\rangle \left\langle x^{\prime }\right\vert dxdx^{\prime }+\int \rho
_{ND}(x,x^{\prime })\left\vert x\right\rangle \left\langle x^{\prime
}\right\vert dxdx^{\prime }  \label{delta3}
\end{equation}%
Taking the trace of $\rho $ we obtain%
\begin{equation}
Tr(\rho )=\int \left\langle x^{\prime \prime }\right\vert \rho \left\vert
x^{\prime \prime }\right\rangle dx^{\prime \prime }=\frac{1}{\pi }\underset{%
\epsilon \rightarrow 0}{\lim }\int \rho _{D}(x)\frac{\epsilon }{%
(x-x^{\prime })^{2}+\epsilon ^{2}}\delta (x^{\prime }-x)dxdx^{\prime }+\int
\rho _{ND}(x,x^{\prime })\delta (x^{\prime }-x)dxdx^{\prime }  \label{delta4}
\end{equation}%
We can proceed with the integral of the Dirac delta in both terms, so
finally we obtain%
\begin{equation}
Tr(\rho )=\underset{\epsilon \rightarrow 0}{\lim }\frac{1}{\pi }\frac{1}{%
\epsilon }\rho _{D}+\rho _{ND}  \label{delta5}
\end{equation}%
where%
\begin{equation}
\rho _{D}=\int \rho _{D}(x)dx\text{ \ \ \ \ \ \ }\rho _{ND}=\int \rho
_{ND}(x,x)dx  \label{delta6}
\end{equation}

In the case of the quantum state of eq.(\ref{c9.3}) we will have (we do not
put the $\underset{\epsilon \rightarrow 0}{\lim }$ for simplicity)%
\begin{equation}
Tr(\rho _{int}^{(n,p)})=\underset{i=1}{\overset{L(n,p)}{\prod }}\left( 
\frac{\rho _{D}^{(n,p,i)}}{\pi \epsilon }+\rho _{ND}^{(n,p,i)}\right) =%
\underset{j=-L(n,p)}{\overset{0}{\sum }}\gamma _{j}^{(n,p)}\epsilon ^{j}
\label{delta7}
\end{equation}%
where in particular%
\begin{equation}
\gamma _{0}^{(n,p)}=\overset{L(n,p)}{\underset{i=1}{\prod }}\rho
_{ND}^{(n,p,i)}\text{ , \ ... \ , \ }\gamma _{L(n,p)}^{(n,p)}=\frac{1}{\pi
^{L(n,p)}}\overset{L(n,p)}{\underset{i=1}{\prod }}\rho _{D}^{(n,p,i)}
\label{delta8}
\end{equation}

In \cite{PR1} we suggest the relation between the Dirac delta valuated at
zero and the pole of the dimensional regularization but we do not prove it.%
\footnote{%
From a different point of view, if we expand in Taylor series the
representation of the Dirac delta of eq.(\ref{delta1}) we obtain $\delta (x)=%
\underset{\epsilon \rightarrow 0}{\lim }\frac{1}{\pi }(\frac{1}{\epsilon }-%
\frac{x^{2}}{\epsilon ^{3}}+\frac{x^{4}}{\epsilon ^{5}}+...)$. Taking the
trace of the quantum state implies to replace $x=0$ in the representation of
the Dirac delta.}

\section{Relation between the vacuum energy and the space volume}

To obtain the relation between the energy of the vacuum and the space volume 
$V$ we can recall the renormalized result of eq.(\ref{va15.1}):%
\begin{equation}
\left\vert \left\langle \Omega _{0}\right\vert \Omega \rangle \right\vert
^{2}e^{-iE_{0}2T}=\underset{k=0}{\overset{+\infty }{\sum }}\frac{\left(
2TV\right) ^{k}}{k!}\left( \underset{p=1}{\overset{+\infty }{\sum }}\left(
-i\lambda _{0}\right) ^{p}\beta _{0}^{(0,p)}\right) ^{k}  \label{ener1}
\end{equation}%
then we can call%
\begin{equation}
(-i)^{k}R(k)=\left( \underset{p=1}{\overset{+\infty }{\sum }}\left(
-i\lambda _{0}\right) ^{p}\beta _{0}^{(0,p)}\right) ^{k}  \label{ener2}
\end{equation}%
which implies that%
\begin{equation}
R(k)=\left[ R(1)\right] ^{k}  \label{ener2.1}
\end{equation}%
where%
\begin{equation}
R(1)=\underset{p=1}{\overset{+\infty }{\sum }}\left( -i\lambda _{0}\right)
^{p}\beta _{0}^{(0,p)}  \label{ener2.2}
\end{equation}%
then, eq.(\ref{ener1}) reads%
\begin{equation}
\left\vert \left\langle \Omega _{0}\right\vert \Omega \rangle \right\vert
^{2}e^{-iE_{0}2T}=\underset{k=0}{\overset{+\infty }{\sum }}\frac{1}{k!}%
(-i2TVR(1))^{k}=e^{-i2TVR(1)}  \label{ener4}
\end{equation}%
then the vacuum energy reads%
\begin{equation}
E_{0}=VR(1)-\frac{i}{2T}\ln (\left\vert \left\langle \Omega \right\vert
\Omega _{0}\rangle \right\vert ^{2})  \label{ener5}
\end{equation}%
in particular, for $T\rightarrow \infty $%
\begin{equation}
E_{0}\sim V  \label{ener6}
\end{equation}%
which is the desired result (see \cite{PS}, page 98). This result is valid
if the $R(1)$ as a sum converges. In fact, the ratio test applied to
argument of the sum in eq.(\ref{ener2.2}) implies that%
\begin{equation}
\underset{p\rightarrow \infty }{\lim }\frac{\left\vert \beta
_{0}^{(0,p+1)}\right\vert }{\left\vert \beta _{0}^{(0,p)}\right\vert }<\frac{%
1}{\lambda _{0}}  \label{ener7}
\end{equation}%
This inequality can be tested on the l.h.s. step by step using dimensional
regularization. Is not the purpose of this work to prove the convergence of
the $n=0$ correlation function of $\phi ^{4}$ theory, besides that it would
be a long task.

\section{The projection in algebraic terms}

Let us remember the transformation of eq.(\ref{finite1}). For simplicity we
will describe it when there are only one diagonal state and one non-diagonal
state, in this case, the transformation act in the following way%
\begin{equation}
\Pi (\rho )=\rho -\rho _{D}  \label{pro1}
\end{equation}%
This transformation is linear%
\begin{equation}
\Pi (\rho ^{(1)}+\rho ^{(2)})=\rho ^{(1)}+\rho ^{(2)}-(\rho _{D}^{(1)}+\rho
_{D}^{(2)})=\rho ^{(1)}-\rho _{D}^{(1)}+\rho ^{(2)}-\rho _{D}^{(2)}=\Pi
(\rho ^{(1)})+\Pi (\rho ^{(2)})  \label{pro1.1}
\end{equation}%
Then it is a projector because%
\begin{equation}
\Pi ^{2}(\rho )=\Pi (\Pi (\rho ))=\Pi (\rho -\rho _{D})=\Pi (\rho )-\Pi
(\rho _{D})=\rho -\rho _{D}-(\rho _{D}-\rho _{D})=\Pi (\rho )  \label{pro1.2}
\end{equation}%
or by using that the diagonal part of the transformed state $\Pi (\rho )_{D}$
is zero%
\begin{equation}
\Pi (\Pi (\rho ))=\Pi (\rho )-\Pi (\rho )_{D}=\rho -\rho _{D}-\Pi (\rho
)_{D}=\Pi (\rho )  \label{pro1.3}
\end{equation}%
In this sense, the projector can be written as%
\begin{equation}
\Pi =I-Q  \label{pro2}
\end{equation}%
where%
\begin{equation}
Q(\rho )=\rho -\rho _{ND}  \label{pro3}
\end{equation}%
then, eq.(\ref{pro2}) is the relation of orthogonal projections. In fact%
\begin{equation}
Q\Pi (\rho )=Q(\rho -\rho _{D})=Q(\rho )-Q(\rho _{D})=\rho _{D}+\rho
_{ND}-\rho _{ND}-\rho _{D}=0  \label{pro5}
\end{equation}%
which implies that $\Pi (\rho )$ is the null space of $Q$.

What the projector does is to subtract from $\rho $ its diagonal part, which
gives a divergent structure when we compute the trace with the observable.
In this sense, to subtract the $\epsilon ^{-l}$ terms via a projection is
similar to the minimal subtraction, where an operator $K$ is defined to pick
out the pure poles terms of the dimensional regularization (see \cite%
{critical}, eq. 9.76)):%
\begin{equation}
K\left[ \underset{n=-k}{\overset{+\infty }{\sum }}A_{n}\epsilon ^{n}\right] =%
\underset{n=-k}{\overset{-1}{\sum }}A_{n}\epsilon ^{n}  \label{pro6}
\end{equation}%
then%
\begin{equation}
\left( I-K\right) \left[ \underset{n=-k}{\overset{+\infty }{\sum }}%
A_{n}\epsilon ^{n}\right] =\underset{n=0}{\overset{+\infty }{\sum }}%
A_{n}\epsilon ^{n}=A_{0}+A_{1}\epsilon +...  \label{pro7}
\end{equation}%
In fact, $K^{2}=K\,$, then $K$ is a projector. The main difference is that
our projector acts on a quantum state and not over a Laurent series. It will
be source of future works to study the relationship between the projection
procedure and the $BPHZ$ subtraction method \cite{ref}.

Finally, we can rewrite the projector that acts on the whole Liouville space
in algebraic language. For this, in the order $p$ of the perturbation
expansion we have the following Hilbert spaces:%
\begin{equation}
\mathcal{H}^{(n,p)}=\mathcal{H}_{ext}\underset{i=0}{\overset{L(n,p)}{\oplus }%
}\mathcal{H}^{(i)}  \label{sys1}
\end{equation}%
The total Hilbert space to all orders in the perturbation theory reads%
\begin{equation}
\mathcal{H}=\mathcal{H}^{(n,0)}\oplus \mathcal{H}^{(n,1)}\oplus ...\oplus 
\mathcal{H}^{(n,p)}=\underset{i=0}{\overset{p}{\oplus }}\mathcal{H}^{(n,p)}
\label{sys2}
\end{equation}%
The observables are defined in the Liouville space $\mathcal{N}$:%
\begin{equation}
\mathcal{N=H\otimes H=}(\underset{i=0}{\overset{p}{\oplus }}\mathcal{H}%
^{(n,p)})\otimes (\underset{i=0}{\overset{p}{\oplus }}\mathcal{H}^{(n,p)})=%
\underset{i=0}{\overset{p}{\oplus }}\mathcal{N}^{(i)}  \label{sys3}
\end{equation}%
We can decompose as (see eq.\ref{k})):%
\begin{equation}
\mathcal{\mathcal{\mathcal{N}}}_{vh}\mathcal{=N}_{S}\oplus \mathcal{N}%
_{R}\subset \mathcal{\mathcal{\mathcal{N}}}  \label{k}
\end{equation}%
Then, the relevant Liouville space will reads%
\begin{equation}
\mathcal{\mathcal{\mathcal{N}}}_{vh}=\underset{i=0}{\overset{p}{\oplus }}%
\left( \mathcal{N}_{S}^{(i)}\oplus \mathcal{N}_{R}^{(i)}\right)  \label{k1}
\end{equation}%
Because the states must be considered as linear functionals over the space $%
\mathcal{\mathcal{\mathcal{N}}}_{vh}$ ($\mathcal{\mathcal{\mathcal{N}}}%
_{vh}^{\prime }$ the dual of space $\mathcal{\mathcal{\mathcal{N}}}_{vh}$):%
\begin{equation}
\mathcal{N}_{vh}^{\prime }\mathcal{=}\underset{i=0}{\overset{p}{\oplus }}%
\left( \mathcal{N}_{S}^{\prime (i)}\oplus \mathcal{N}_{R}^{\prime (i)}\right)
\label{m}
\end{equation}%
Then, the projector will be a map from $\mathcal{N}_{vh}^{\prime }$ to $%
\mathcal{N}_{R}^{\prime }$:%
\begin{equation}
\Pi =\Pi ^{(0)}\oplus ...\oplus \Pi ^{(p)}:\mathcal{N}_{vhS}^{\prime
}\rightarrow \mathcal{N}_{R}^{\prime }  \label{proy}
\end{equation}

\textit{This is the simple trick that allows us to neglect the singularities
(i.e. the }$\delta (x-x^{\prime }))$ \textit{in a rigorous mathematical way
and to obtain correct physical results}. Essentially we have defined a new
dual space $\mathcal{N}_{vh}^{\prime }$ (that contains the states $\rho $
without divergences) that are adapted to solve our problem.

So, essentially we have substituted an \textquotedblleft ad
hoc\textquotedblright\ counterterm procedure (or an ad hoc subtraction
procedure \cite{ref}) with a clear physical motivated theory.\textit{\ }%
These are the essential features of the proposed formalism, where the deltas
are absent.

\end{document}